\documentclass[aps,prl,twocolumn,psfig,epsfig,showpacs,preprintnumbers,amsmath,amssym,superscriptaddress]{revtex4}
\usepackage{graphicx}
\usepackage{bm}
\usepackage{epsf}
\begin{document} 
\def\kv{ {\bf k}}
\def\xv{ {\bf x}}    

\newlength{\pcm}
\setlength{\pcm}{0.5cm}
\newlength{\pmm}
\setlength{\pmm}{0.1\pcm}
\newcommand{\kk}{\hspace{+3.0\pmm}}
\newcommand{\nn}{\nonumber}

\newcommand {\vertex}{\,\epsfxsize=2.5\pcm \parbox{1.5\pcm}{\epsfbox{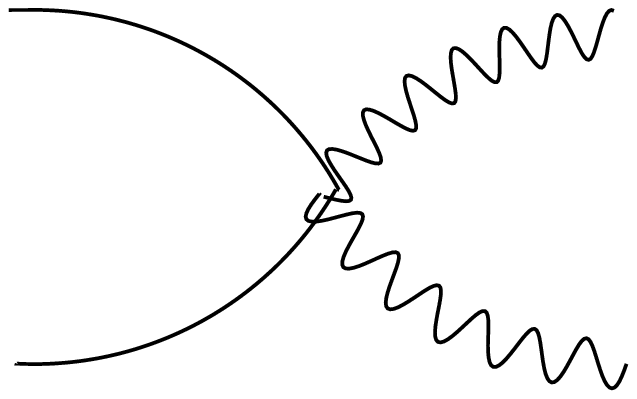}}\,}

\title{Absorbing state phase transitions with a non-accessible vacuum}
\author{Omar Al Hammal}
\author{Juan A. Bonachela}
\author{Miguel A. Mu\~noz}
\affiliation{Instituto~de~F\'\i sica~Te\'orica~y~Computacional~Carlos~I  and \\
Departamento de Electromagnetismo y ~F\'\i sica de la Materia, \\
~Facultad~de~Ciencias,~Universidad~de~Granada,~18071~Granada,~Spain}
\date{\today}

\begin{abstract}
We analyze from the renormalization group perspective a universality
class of reaction-diffusion systems with absorbing states. In this
class, models where the vacuum state is not accessible are represented
as the set of reactions $2 A \rightarrow A$ together with creation
processes of the form $A \rightarrow n A$ with $n \geq 2$.  This class
includes the (exactly solvable in one-dimension) {\it reversible}
model $2 A
\leftrightarrow A$ as a particular example, as well as many other {\it
non-reversible} sets of reactions, proving that reversibility is not
the main feature of this class as previously thought. By using field
theoretical techniques we show that the critical point appears at zero
creation-rate (in accordance with known results for the reversible
case) and it is controlled by the well known pair-coagulation
renormalization group fixed point, with non-trivial exactly computable
critical exponents in any dimension. Finally, we report on Monte-Carlo
simulations, confirming the field theoretical predictions in one and
two dimensions for various reversible and non-reversible sets of
reactions.
\end{abstract}

\maketitle

\section{Introduction}
In a recent paper Elgart and Kamenev \cite{EK} have proposed a
classification of absorbing state phase transitions, a subject that
has been one of the central pillars of non-equilibrium statistical
mechanics over the last decade \cite{MD,Reviews_AS}.  The strategy
they follow is elegant and powerful. The main idea is to (i) write
down using standard techniques the generating functional (or,
equivalently, the effective Hamiltonian) for a given
reaction-diffusion system; (ii) inspect the phase space in
saddle-point approximation paying special attention to the
``zero-energy'' manifolds which determine the topological properties;
(iii) detect possible structural changes in the phase portrait: the
birthmark of phase transitions, and (iv) classify them according to
basic topological properties. This procedure is a natural extension to
non-equilibrium problems of the rearrangement of
thermodynamic-potential minima occurring at equilibrium phase
transitions. Hence, it allows for a categorization of universality
classes attending to symmetry principles, conservation laws, and few
other relevant ingredients, which determine the phase-space topology
and its possible structural changes. Establishing the limits of
validity of the saddle-point approximation within this context and
developing systematic improvements to it remain as fundamental open
problems.

Using this strategy, Elgart and Kamenev report on $5$ non-trivial
universality classes with absorbing states, occurring in
one-dimensional systems with just one type of particle
\cite{4,2species}. The first $4$ ones are: (i) {\it directed percolation} (DP)
characterizing generic systems with an absorbing phase transition and
without extra symmetries, conservation laws, quenched disorder, nor
long-range interactions \cite{Reviews_AS,conjecture}, (ii) the usually
called {\it parity conserved} (PC) \cite{PC} also known as DP2 or
generalized voter class \cite{GV} which includes two symmetric
absorbing states, (iii) the very elusive {\it
pair-contact-process-with diffusion} (PCPD) class in which all
reactions involve pairs of particles \cite{Review_PCPD,support}, and
(iv) the {\it triplet-contact-process-with diffusion} (TCPD) in which
reactions involve triplets of particles \cite{TCPD}.

In this paper we focus on the fifth class in \cite{EK}.  It describes
the reversible reactions $ A \rightarrow 2 A$ and $2A
\rightarrow A$ occurring at rates $\mu$ and $\sigma$
respectively. This model was solved exactly in one dimension more than
twenty years ago in a seminal paper by Burschka, Doering, and
ben-Avraham \cite{Exact} by employing the {\it empty interval method}
\cite{Empty}. Finite-size properties, scaling functions, and critical
exponents have also been exactly computed for this reversible model
and for variations of it
\cite{Exact} in one dimension. Note that except for the absence of
one-particle spontaneous annihilation, $A\rightarrow 0$, this set of
reactions coincides with the contact process \cite{MD} a well-known
model in the robust DP class \cite{Reviews_AS}. It is, therefore,
interesting to elucidate which is the main relevant difference in the
renormalization group sense, giving rise to a non-trivial non-DP type
of scaling. From considerations in \cite{EK} it seems that the fact
that the reactions are reversible plays such a relevant role, but as
we will illustrate, {\it reversibility is a sufficient, but not a
necessary, condition}.

From the field theoretical point of view, Cardy and T\"auber had
obtained in their seminal article \cite{CT} a one-loop calculation of
critical exponents for the closely related set of reactions $2 A
\leftrightarrow 0$, while in a recent paper Jack, Mayer, and Sollich
have shown that such one-loop results are also valid for $2A
\leftrightarrow A$ and have to be exact owing to the existence of
{\it detailed balance} for reversible reactions
\cite{Jack}. Therefore, two or more loop corrections should cancel
out, even if this is not explicitely shown in \cite{Jack}. In any
case, the main results are that the critical point is located at
$\mu_c=0$ (any non-vanishing branching rate leads to sustained
activity) and the order-parameter critical exponent is $\beta=1$.  The
long-time long-distance properties turn out to be controlled by the
well-known ``pure'' pair-coagulation ($2A \rightarrow A$) RG fixed
point \cite{Peliti,CT,RD,Jack} and all exponents can be computed in
any dimension.

In this paper, we perform a full diagrammatic expansion of various
reaction-diffusion models extending previous analyses to all orders in
perturbation theory. First, we recover the previously known results
for the reversible model $2A \leftrightarrow A$. Afterward, using the
intuition developed from the previous full diagrammatic analysis we
construct different sets of {\it non-reversible} reactions, and argue
that they belong to this same universality class. Its key ingredient
turns out to be the absence of an accessible vacuum state, i.e. there
is no reaction $m A \rightarrow 0$ but just pair-coagulation, combined
with creation reactions of the form $A \rightarrow n A$. The {\it
reversible reaction}, $n=2$ discussed in \cite{EK} and \cite{Jack} is
just a representative of this broader class: reversibility (which
tantamount to the detailed-balance condition in \cite{Jack}) is a
sufficient but not a necessary requirement.

Let us remark that reactions as $2A \leftrightarrow 0$ and its
non-reversible extensions $2A \rightarrow 0$, $0 \rightarrow 3A, ~ 4A,
...$ can also be argued to belong to this same class. In these cases,
the vacuum state is accessible, but it is not stable, so they are not
genuine absorbing state models.

To verify the field-theoretical predictions we perform Monte-Carlo
simulations for various non-reversible sets of reactions, implemented
with and without hard-core exclusion (``fermionic'' or ``bosonic'',
respectively) in one and two dimensions. All critical exponents, are
in perfect agreement with the RG predictions, confirming the existence
of a robust universality class, broader than thought before.

Before proceeding, we should underline that while many of the results
contained in this paper are already known (some from exact solutions
of the reversible model in one dimension \cite{Exact} and some from
similar perturbative calculations combined with symmetry arguments
\cite{EK,Peliti,CT,RD,Jack}), a systematic presentation of them,
focusing the attention on universality aspects is, to the best of our
knowledge, lacking in the literature. This paper aims at filling this
empty space and at providing a comprehensive picture of this
universality class, extending it to non-reversible reactions without
an accessible vacuum state.

\section{Field theory analysis of $ 2A \leftrightarrow A$}

The techniques employed in this section are standard and we refer the
reader to \cite{GF,ZJ} and more specifically to \cite{Peliti,CT,RD}
for more detailed calculations and/or pedagogical presentations.

Let us apply the Doi-Peliti formalism \cite{GF,CT,RD} (see also
\cite{Poisson,Gardiner}) to the reversible set of reactions $ A \rightarrow 2A$
and $ 2A \rightarrow A$ occurring at rates $\mu$ and $\sigma$
respectively.  They can be cast into a generating functional whose
associated (bosonic) action is
\begin{eqnarray} 
 {\cal S} [ \phi,\pi ] & = & \int dt \int d^dx [ \pi ( \partial_t \phi
 - D \nabla^2 \phi ) - H[\phi,\pi] ], ~~ with \nonumber \\ H[\phi,\pi]
 & = &(\pi^2 -\pi) (\mu \phi - \sigma \phi^2 ),
\label{action}
\end{eqnarray}
where $\phi(\xv, t)$ and $\pi(\xv,t)$ are the density and the response
fields respectively (some spatial and time dependences have been
omitted for simplicity). For a general process $k A \rightarrow j A$
with $k$ and $j$ integer numbers, the associated effective Hamiltonian
in this formalism includes a factor $[\pi^j - \pi^k] \phi^k$, which
{\it is proportional to $[\pi^2 -\pi]$ if and only if the absorbing
state is not accessible, i.e. $j> 0$ and $k>0$}.

For readers with more intuition in terms of stochastic equations, an
associated Langevin equation can be easily derived:
\begin{equation}
\partial_t \phi(\xv,t)=  D \nabla^2 \phi+ \mu \phi -\sigma \phi^2 +
\sqrt{\mu \phi -\sigma \phi^2} \eta(\xv,t)
\label{langevin}
\end{equation}
where $\eta(\xv,t)$ is a Gaussian white noise. Let us emphasize the
similarity between Eq.(\ref{langevin}) and the Langevin equation for
the DP class \cite{conjecture,Reviews_AS}.  Despite of this alikeness,
Eq.(\ref{langevin}) is not free from interpretation difficulties as
the density field is not a real-valued one, but develops an imaginary
part \cite{Gardiner}.  For this reason we avoid using it and center
the forthcoming discussion on Eq.(\ref{action}).

Owing to the fact that the effective Hamiltonian in Eq.(\ref{action}),
$H[\phi,\pi]$, is proportional to $(\pi^2-\pi)$, $\pi=0$ and $\pi=1$
are zero-energy manifolds. The existence of these two constant-$\pi$
solutions is, according to \cite{EK}, at the basis of the non-DP
behavior of this model.  Indeed, the four zero-energy solutions:
$\pi=0$, $\pi=1$, $\phi=0$, and $\phi=\mu/\sigma$, define a
rectangular geometry in the phase portrait (see Fig. 1b and
\cite{EK}), which should be compared with the standard, triangular, DP
topology, for which only one constant-$\pi$ solution exists (see
Fig. 1a and \cite{EK}) as we illustrate now.

\begin{figure}
\includegraphics[width=76mm,clip]{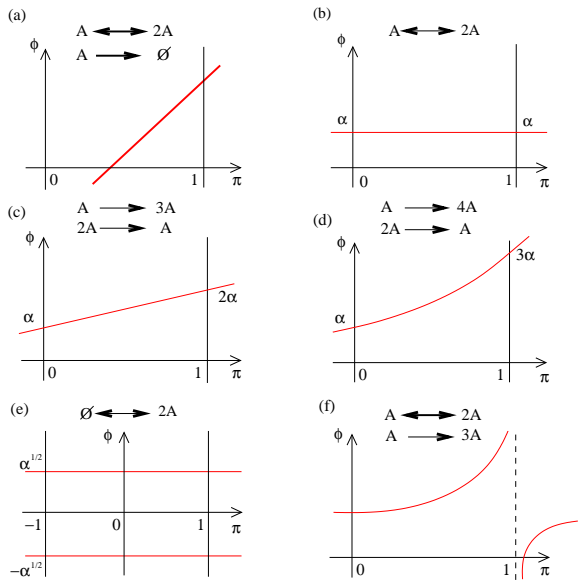}
\caption{Schematic zero-energy manifolds (bold lines) for different reactions. 
The line marked in red (non-trivial manifold, depending on the control
parameter $\mu$) moves downward upon approaching the critical point in
all cases: $\alpha=\mu/\sigma$. While directed percolation is
characterized by a triangular structure as in (a), models without an
accessible vacuum state have a different form, being rectangular (b),
trapezoidal (c), or more complicated geometries (d), for different
reactions. For reversible models with an unstable vacuum,  $\phi=0$ is
not a zero-energy manifold (e). The structures in (c) and (d) are not
robust under RG flow, but evolve to non-closed topologies as the one
in figure (f).}
\label{diagrams}
\end{figure}

It is worth noticing that the common factor $(\pi^2-\pi)$ in
Eq.(\ref{action}), arising from the fact that the absorbing state is
not accessible, can be interpreted as a {\it subtle symmetry} between
all the coefficients of (noise) terms proportional to $\pi^2$ and
their corresponding (deterministic) ones proportional to $-\pi$.
Indeed, it is closely related to the detailed-balance symmetry
discussed in \cite{Jack}. If an additional reaction $A \rightarrow 0$
occurring at rate $\lambda$ is switched on, a term $\lambda \phi
(1-\pi)$ has to be added to the Hamiltonian. In such a case,
$(\pi^2-\pi)$ is not a common factor, the subtle symmetry is broken
and $\pi=0$ is not a zero-energy solution anymore. This leads to the
triangular topology for zero-energy manifolds in the phase space (Fig
1a) and, hence, to DP-scaling. Something similar occurs by switching
on any other reaction as $m A \rightarrow 0$, with $m \geq 2$,
converting the vacuum into an accessible state.

Let us present a different argument leading to the same
conclusion. From standard na\"ive power counting and relevance
arguments one could be tempted to conclude that this problem is in the
DP class, and that the critical dimension is $d_c=4$. Indeed, as said
before, the leading (lowest order) terms in both the deterministic
part and the noise are identical to their analogous ones in the DP
field theory \cite{conjecture,Reviews_AS}, for which $d_c=4$. The only
way out of this na\"ive (and wrong) conclusion is that, at the
critical point where the linear-deterministic term coefficient
vanishes, the coefficient of the leading DP-like noise term $\pi^2
\phi$ also vanishes owing to the abovementioned subtle symmetry.
This opens the door for higher order noise terms to control the
(non-DP) scaling. Indeed, a proper power counting analysis reveals
that, as the interaction Hamiltonian is proportional to $(\pi^2 -
\pi)$, $\pi$ has to be dimensionless, which leads to $[\phi]=
\Lambda^d$ (where $\Lambda$ has dimensions of momentum) to ensure a
dimensionless action, and consequently to $[\mu] =\Lambda^2$ and
$[\sigma]= \Lambda^{2-d}$. Therefore, the theory upper critical
dimension is $d_c=2$ \cite{CT,Jack,EK}.

The existence of the common factor $(\pi^2-\pi)$ in Eq.(\ref{action})
implies that the $\mu$-dependent non-trivial manifold and the trivial
one, $\phi=0$, merge at the critical point rather than intersecting in
just a point as in DP. This is the key reason for the models without
an accessible vacuum to exhibit a different type of scaling.  To
substantiate this assertion we need to prove that the previous
bare-action symmetry, or associated topological structure, survives to
the inclusion of fluctuations (i.e. it remains valid beyond mean-field
\cite{MF}).

The basic elements for a complete perturbative expansion at a
diagrammatic level are: the usual DP-noise vertex $(\pi^2 \phi)$
\cite{conjecture,CT,RD}, the pair-coagulation ones ($\pi \phi^2$ and
$\pi^2 \phi^2$), as well as the propagator ${ (-i \omega + D \kappa^2
+ \mu)}^{-1}$ \cite{Peliti,CT,RD}.  Diagrammatically, we represent
response fields by wavy lines and the density fields by straight
ones. For instance, the pair-coagulation noise vertex is depicted as
$~~~\sigma^2\vertex ~~~~~$
and analogous figures, with different numbers of straight and/or wavy
lines are employed for the other vertices and the propagator.

In order to perform a sound perturbative expansion to all orders in
perturbation theory, we choose to write separately diagrams with and
without corrections proportional to $\mu$. The second group includes
only diagrams with vertices proportional to $\sigma$ (i.e. expansions
of Z-functions in powers of $\sigma$). Simple inspection reveals that
such diagrammatic corrections are those of the pair-coagulation
process, a theory well-known to be {\it super-renormalizable}, i.e.,
all these diagrams can be computed and resummed to all orders. Indeed,
the only possible diagrammatic corrections to the pair-coagulation
vertices, proportional to $\sigma$, have the typical ``bubble''
structure, leading to a geometric series
\begin{eqnarray}
\sigma_R  \vertex \kk \kk & =&~~ \sigma ~~ \vertex    \nn \\               
      & - & ~\sigma^2 ~~\vertex \kk~~ \vertex \nn \\
      & + &~ \sigma^3~~ \vertex\kk~~ \vertex\kk~~  \vertex ~~
        \kk \kk + \ldots \nn \\
       &=& {\sigma \vertex \over  1 + \sigma  \vertex\kk ~~\vertex} \nn,
\label{series}
\end{eqnarray}
where $\sigma_R$ is the renormalized (or ``dressed'') coagulation
coefficient. Omitting external legs in Eq.(\ref{series}), $\sigma_R =
{\sigma \over 1 + \sigma I}= \sigma (1 - \sigma \Sigma)$ with $\Sigma=
{I \over 1 + \sigma I}$, where I denotes the one-loop diagram
evaluated at zero external frequency and arbitrary momentum scale
$\Lambda$ \cite{GF,ZJ,CT}:
\begin{equation}
I = {1 \over (2 \pi)^{(d+1)}} \int d^d \kappa~ d \omega~ {1 \over i
\omega + D \kappa^2} ~ {1 \over -i \omega + D \kappa^2} \propto
{\Lambda^{-\epsilon} \over \epsilon}
\end{equation}
with $\epsilon=2-d$.
Similar expressions are obtained for all the renormalized coefficients
by just changing the leftmost and/or the rightmost vertex of the
series in Eq.(\ref{series}). If these were the only corrections,
(i.e. if the RG fixed point was at $\mu_c=0$ so that diagrams
including corrections proportional to $\mu$ would not give any
non-vanishing contribution) then the renormalized parameters would be:
\begin{eqnarray}
&& \mu_R= Z_\mu \mu = \mu ( 1 - \sigma \Sigma) \nonumber \\ &&
\sigma_R= Z_\sigma \sigma = \sigma (1 - \sigma \Sigma ).
\label{zeta}
\end{eqnarray}
for the two coefficients proportional to $\mu$ and the two
proportional to $\sigma$ respectively, showing that the subtle
symmetry is not broken.  The corresponding flow equations would be
\begin{eqnarray}
&& \partial_l \mu_R = \mu (2 - \sigma \partial_l \Sigma) 
\nonumber \\ && \partial_l \sigma_R = \sigma (\epsilon -
\sigma \partial_l \Sigma)
\label{flow}
\end{eqnarray}
where $\partial_l$ stands for the logarithmic derivative with respect
to the momentum scale at which integral are evaluated
\cite{GF,CT,ZJ,RD}. 
For $\epsilon <0$, i.e. $d>2$, the trivial (mean field) solution
$\sigma=\mu=0$ is infrared stable, while for $\epsilon >0$, the only
infrared stable fixed point is $\sigma* = \epsilon /(\partial_l
\Sigma)$ with $\mu=0$.  Plugging this into the first equation in 
(\ref{flow}), we obtain the anomalous scaling dimension of $\mu_R$,
$[\mu_R]= 2 - \epsilon = d$, which coincides with the one-loop result
obtained in \cite{CT,EK} (see also \cite{Jack}).

The change in the scaling dimension of the ``mass'' term, from its
na\"ive value $[\mu]=2$ to the renormalized exact one, $[\mu_R]=d$,
induces a change in all critical exponents corresponding to magnitudes
measured away from the critical point with respect to their
corresponding mean field values. Moreover, as happens in
pair-coagulation, there is no further renormalization required for the
fields nor the diffusion constant
\cite{Peliti,CT,norenorm} and therefore all exponents can be exactly
computed at any dimension. For instance, the scaling dimension of the
field is $\Lambda^d$ and, hence, scales as $\mu_R$, implying $\beta=1$
in any dimension. Using the same logic one obtains $\nu_{||}=2$,
$\nu_{\perp} =1$ for the correlation time and correlation length
exponents, while right at the critical point $z=2$. Using standard
scaling relations, the density of particles as a function of time
decays in one dimension with an exponent $\theta=1/2$, while in $d=2$
a similar calculation leads to logarithmic corrections and, in
particular, to a decay $\ln(t)/t$, while $\beta$ remains equal to $1$
\cite{CT}.

In order to prove that the fixed point with $\mu_c=0$ is not just {\it
a solution} but also the {\it only one} one should consider all the
possible diagrams (even if this can be done only in a symbolic form
\cite{note1}), write down the $4$ Z-functions, analogous to eq.(\ref{zeta})
for the $4$ vertex functions ($2$ proportional to $\mu$ and $2$ to
$\sigma$ in their bare form). Doing this, it is straightforward to
check that $3$ different and independent flow equations are
obtained. The fourth one is not independent owing to the usual duality
symmetry \cite{conjecture} but this is not important for the
argumentation here. As there are only $2$ independent bare parameters,
there is no way to find a fixed point for this set of $3$ independent
equations except for the trivial one $\mu_c=0$, which simultaneously
satisfies in a trivial way the first $2$ equations, and leads back to
the preceeding calculation, to the symmetry preserving
Eq.(\ref{zeta}), and to the same set of exponents.

Note that in models in the DP class, where the na\"ive power counting
is different, with $d_c=4$, only $2$ independent parameters in the
flow equations need to be fine tuned to zero. The third one
(corresponding to the highest order noise coefficient) is irrelevant
(flows to zero) already at mean field level and, therefore, does not
require fine tuning to vanish asymptotically. Hence, contrarily to the
previous case, a non-trivial solution, $\mu_c \neq 0$ exists leading
to a DP fixed point.

As pointed out in \cite{Jack}, the reversible reaction studied here
and $2A \leftrightarrow 0$ share the same type of critical
behavior. Indeed, the Hamiltonian in this latter case is $(\pi^2 -1)
(\mu - \sigma \phi^2 )$ where, as before, $\mu$ and $\sigma$ are the
creation and annihilation rate respectively. The zero-energy manifolds
are: $\pi=\pm 1$ and $\phi=\pm \sqrt{\mu/\sigma}$ (fig. 1e). They
define a quadrangular structure, as the one described above, but in
this case $\phi=0$ is not an invariant manifold: the vacuum state is
accessible but it is not stable, so it is not properly an absorbing
state phase transition. A perturbative analysis analogous to the one
above can be done for the present case (indeed this is the model
studied in
\cite{CT,Jack}) and leads to the same set of critical exponents; here
the common factor $\pi^2-1$ plays the role of the subtle symmetry
above. 

Finally, for reversible coagulation reactions involving {\it triplets}
instead of pairs, $A \leftrightarrow 3A$, we obtain similar results:
vanishing critical point and exactly computable exponents, but the
critical dimension is $d_c \leq 1$ in this case \cite{EK}.

\section{Extension to  non-reversible reactions}

A careful but simple inspection of the arguments in the preceeding
section leads to the conclusion that none of the reported results
depends on the fact that the creation reaction is of the form $A
\rightarrow 2A$. As will be argued in this section, most of them 
apply to more general {\it non-reversible} processes with creation
reactions as $A \rightarrow nA$. For these, the creation part of the
Hamiltonian is $ \mu (\pi^n - \pi) \phi$, which together with the
pair-coagulation terms $ \sigma (\pi^2 - \pi) \phi^2$ guarantees that
$(\pi^2-\pi)$ can be extracted as a common factor for non-reversible
bare Hamiltonians, and hence $\pi$ is dimensionless and $\pi=0$ and
$\pi=1$ are zero-energy solutions as in the $n=2$ case. For example
for $n=3$, $H=\mu (\pi^3-\pi) \phi -\sigma(\pi^2 - \pi) \phi^2 =
(\pi^2-\pi) [\mu (\pi+1) \phi -\sigma \phi^2]$. The existence of such
a common factor in the {\it bare} Hamiltonian is, as explained before,
guaranteed if and only if the vacuum state is not accessible.

For the family of non-reversible models with $n>2$, the geometry of
the zero-energy manifolds of the bare Hamiltonian {\it is not a
rectangular one} as occurs for the reversible set of reactions with
$n=2$ \cite{EK}. For instance, for $n=3$ one obtains a {\it
trapezoidal geometry} (zero-energy solutions: $\pi=0$, $\pi=1$,
$\phi=0$ and $\phi= (\pi +1) \mu/\sigma$, (see figure 1c), but the
overall {\it topology} is not changed.  Indeed, as the critical point
is approached the difference between the rectangle and the trapezium
becomes negligible, and at criticality this manifold merges with the
$\phi=0$ one. Analogously, for $n=4$ one obtains a quadrangle with $3$
straight lines and a curved one ($\phi=(\pi^2+\pi+1)\mu/\sigma$) (see
fig. 1d), which also becomes closer and closer to the horizontal line
upon approaching the critical point.  In all cases, the non-trivial
$\mu$-dependent manifold merges with the absorbing-state one $\phi=0$
at the critical point, and this constitutes the main trait of this
class as will be illustrated here: in DP they intersect at criticality
at a single point, in PC they intersect in one point in the active
phase and merge at criticality
\cite{EK}, while in the class under scrutiny, they do not
intersect in the active phase and merge at the critical point.

Note that, as $\pi$ is dimensionless, all the different processes for
different values of $n$ are equally relevant at mean-field level (they
just differ in powers of $\pi$). As a consequence, the na\"ive scaling
dimensions for any $n>2$ are as in the preceeding section, leading to
$d_c=2$.  It is also important to realize that {\it higher-order
processes generate effectively lower-order ones} (in particular, $A
\rightarrow 2A$ is always generated) and all of them share the same
degree of {\it na\"ive} relevancy.  The generation of lower-order
processes induces changes in the zero-energy manifolds, and leads to
combinations of the previous ``pure'' topologies obtained for creation
processes involving only one value of $n$. In order to render the
theory renormalizable, lower order terms have to be included in the
bare Hamiltonian, with coefficients proportional to $\mu$ (as they
have to vanish as $\mu \rightarrow 0$) that we call $\mu_n$. Indeed,
from now on we study physical processes where various types of
creation events with different values of $n$ are simultaneously
present (in particular $n=2$ is always generated).

At a perturbative level, one can proceed as before, and separate
corrections proportional and not proportional to $\mu$. The first
notorious difference with the reversible case is that upon
renormalizing, the shape of some zero-energy manifolds is deformed if
terms with $n \geq 3$ are present. Indeed, owing to the fact that the
coefficients of $\pi^n$, with $n \geq3$, in these generalized processes
renormalize as
\begin{equation}
 \mu'_{n,R}= \mu_n \left( 1 - {n(n-1) \over 2} \sigma \Sigma \right)
\label{3}
\end{equation}
up to one loop \cite{2loop} while the corresponding ``mass''
coefficient renormalizes as in Eq.(\ref{zeta}), different corrections
are generated for these two coefficients equal at a bare level
(therefore, the need to use different names, $\mu'_{n,R}$ and
$\mu_{n,R}=\mu_R$, for the two of them, as a generalization of the
single equation for $\mu_R$ in Eq.(\ref{zeta})). Eq.(\ref{3}) shows
that the scaling dimensions of the non-linear term coefficients,
$\mu'_{n,R}$ varies with $n$: the lower the value of $n$, the more
relevant the corresponding non-linear term.

Proceeding as before, it is straightforward to see by performing a
perturbative expansion around $d_c=2$ that the only way to find a
solution of the RG flow equations at any arbitrary order in
perturbation theory is by fixing $\mu=0$. For instance, considering
creation reactions with $n=2$ and $n=3$, one has $3$ independent
parameters: $\sigma$, $\mu_2$ and $\mu_3$ and $5$ independent flow
equations.  Hence, at criticality all creation rates have to vanish,
and one recovers the fixed point and exponents in the previous
section, so {\it the universality class is preserved under the
introduction of non-reversible reactions}. Note that, in order to
extend the calculation in the preceeding section, it has been enough
to impose that all creation terms are proportional to $\mu$. This
ensures that all of them vanish at the critical point and generate no
extra diagrammatic correction, but they do not need to be all equal as
happens in the reversible case.

We should also emphasize that, as said before, the mass-like terms
associated to each $n$-creation process are all equally relevant and
they all renormalize as $\mu_2$, while the $\mu'_n$ renormalize
differently for $n \geq 3$ (see Eq.(\ref{3})) and hence, $\pi^2-\pi$
is not a common factor of the {\it renormalized} Hamiltonian, except
at the critical point $\mu=0$ where such a subtle symmetry is
restored.  The common factor or subtle symmetry invoked all along the
calculation in the previous section, equivalent to the existence of
reversibility or detailed balance, is not a necessary condition. As a
consequence, the zero-energy manifold structure is affected: the
topology shown in figures 1c and 1d is not stable under the RG flow,
$\pi=1$ is not a zero-energy manifold of the renormalized Hamiltonian,
and the phase portrait structure becomes more complicated (see
Fig. 1f).

Despite of this, we observe that from the phase-portrait point of
view, {\it a key ingredient, not altered upon introducing
non-reversible reactions, is the fact that the non-trivial
$\mu$-dependent manifold and the trivial one $\phi=0$ do not intersect
in the active phase and merge into a degenerate manifold at the
critical point}.  Therefore, the main ingredient of this universality
class is {\it not} the reversibility nor the existence of a common
factor in the renormalized Hamiltonian but the way in which the
non-trivial manifold and the trivial one merge \cite{nnn}. In summary,
{\it reversibility is a sufficient but not a necessary 
requirement}.

For completeness' sake let us comment on another family of reactions
without an accessible vacuum, including higher-order creation
reactions as $k A \rightarrow (k+n) A$ with $k \geq 2$ which exhibit a
different type of scaling behavior. These have to be complemented with
higher order annihilation reactions as $j A \rightarrow l A$ with $j
\geq k$ and $j > l \geq 1$ in order to ensure the existence of a
bounded stationary state.  For instance, taking $2A \rightarrow 3A$
(with rate $\mu$) as a creation reaction together with $2A \rightarrow
A$ (rate $\sigma$), we need another annihilation reaction, as $3A
\rightarrow 2A$ (with rate $\lambda> \mu$) to have a well defined
stationary state. For this case, even if $\pi=0$ and $\pi=1$ are
constant energy solutions (at least at a bare level) the manifold
$\phi=0$ is degenerated, and the non-trivial zero-energy solution,
$\phi=\mu/\lambda - \sigma /(\lambda
\pi)$, intersects the line $\phi=0$ at $\sigma/\mu$ and becomes singular 
at $\pi=0$, originating a very different topology from the one
above. This topology corresponds to the PCPD class
\cite{EK,Review_PCPD}. Therefore, {\it creation from pairs} in systems
without an accessible vacuum leads to a different universality class.

Finally, for non-reversible {\it coagulation reactions involving
triplets} ($3A \rightarrow A$ and $A \rightarrow n A$) we obtain again
that the universality class remains unchanged with respect to the
corresponding reversible reaction (see last paragraph of the previous
section).

\section{MONTE-CARLO SIMULATIONS}
\begin{figure}
\includegraphics[width=76mm,clip]{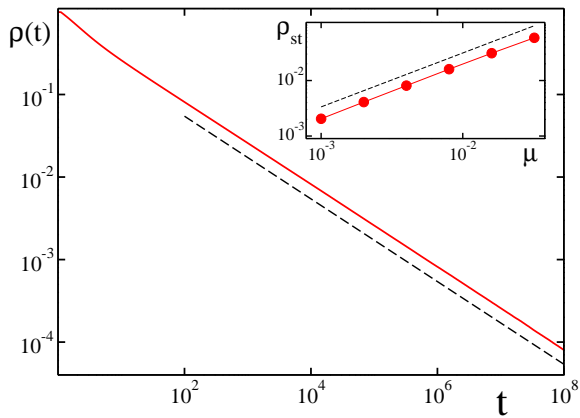}
\caption{(Color online) Results of Monte-Carlo simulations 
for $2A \rightarrow A$ and $A \rightarrow 3A$ implemented in a bosonic
way in one dimension. The decay of the order parameter at criticality
($\mu_c=0$) is given by $t^{-0.5}$ (main plot). This result is well
known as at $\mu=0$ this model coincides with pair-coagulation. The
order-parameter critical exponent is perfectly fitted by $\beta=1.00$
(inset). Very similar results are obtained for the reversible case,
$n=2$, as well as for higher-order non-reversible cases, as $n=4$ and
$n=5$.}
\label{fig1}
\end{figure}
In order to verify the above field theoretical predictions we have
performed Monte-Carlo simulations of the reversible reactions
(reproducing some existing results \cite{Exact}) and, more relevantly,
{\it non-reversible} set of reactions: $2A \rightarrow A$ together
with $A \rightarrow n A$ with $n=3,~4,~ 5$. We have considered two
different implementations: a {\it bosonic} one in which the number of
particles at every site in a lattice is unrestricted (which is the one
directly related to the bosonic field theory presented here), and a
{\it fermionic} one with number occupancy restricted to be $0$ or
$1$. For both of them the same type of numerical experiments have been
performed. Figure 2 shows our main results for $n=3$ in the bosonic
implementation. In the main body, we plot the time evolution of the
order-parameter as a function of time for a one-dimensional lattice of
size $2^{20}$. A clean power-law decay is observed at $\mu_c=0$ with
slope $\theta = 0.500(1)$ in a log-log plot. This is not surprising as
at $\mu=0$ this model coincides with pair-coagulation. Direct
measurements of the order-parameter as a function of the distance to
the critical point (upper inset) lead to $\beta=1.00(1)$. Also, from
measures of the mean-squared distance associated with two point
correlation functions \cite{Jack} one can easily measure $z=2$ in all
the cases under consideration. All the remaining exponents can be
derived using standard scaling laws, providing a full check of the
theoretical predictions for the bosonic model. For the fermionic model
we obtain identical conclusions. In $d=2$ mean-field exponents with
logarithmic corrections have been measured confirming that $d_c=2$
(see figure 3).  In $d=3$, Jack et al. \cite{Jack} showed by means of
Monte-Carlo simulations that the scaling is Gaussian as expected.

\begin{figure}
\includegraphics[width=76mm,clip]{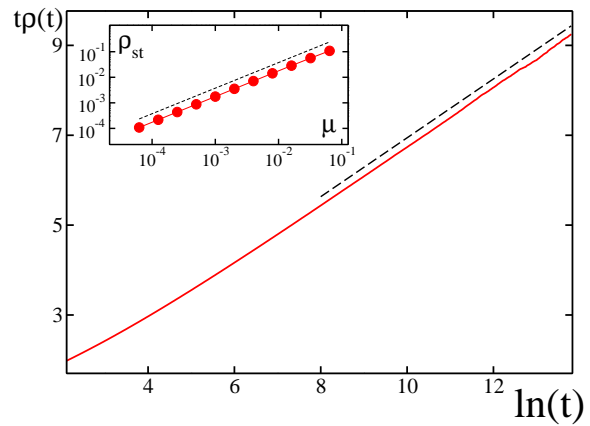}
\caption{(Color online) Results of Monte-Carlo simulations 
for $2A \rightarrow A$ and $A \rightarrow 3A$ implemented in a bosonic
way in two dimensions. The decay of the order parameter at criticality
($\mu_c=0$) is proportional to $\ln(t)/t$ (main plot).  At its
critical point $\mu=0$, this model coincides with pair-coagulation,
for which this is a well known results. The order-parameter critical
exponent is perfectly fitted by $\beta=1.00$ (inset).}
\label{fig2}
\end{figure}
We have also verified that for the sets of non-reversible reactions
with $n=4$ and with $5$ one obtains the same set of critical
exponents, supporting again the theoretical conclusions.

\section{SUMMARY}

We have shown using field theoretical arguments and verified by means
of Monte-Carlo simulations that all reaction-diffusion processes
including pair coagulation $2A \rightarrow A$ and creation in the form
$A \rightarrow nA$ belong to the same universality class, regardless
of whether the reactions are reversible or not. The critical point is
located at zero creation rate, and all critical exponents are
controlled by the well-known pair-coagulation renormalization group
fixed point and can be exactly computed. These conclusions are in
agreement with exactly known results for the reversible model in one
dimension \cite{Exact}. The main ingredient of this class of
absorbing-state transitions is that the vacuum state is not accessible
and creation occurs from individual particles. If creation occurs only
from pairs then scaling is as in the PCPD class while, as soon as a
reaction making the vacuum accessible, as, for example, $2A
\rightarrow 0$ is switched-on, the system recovers standard DP scaling.
There are also models in this universality class as $2A
\leftrightarrow 0$ where the vacuum state is accessible but in these
cases it is not stable: $0 \rightarrow 2A,~ 3A, ...$, so they are not
properly absorbing-state transitions.

We have shown that the topology of the zero-energy manifolds is very
important to unveil universality classes, but there could be many
subtleties leading to surprises. We hope that this work fosters new
studies to clarify some of the still-standing problems on universality
in non-equilibrium critical phenomena.

\vspace{0.35cm}
{\it --} We thank A. Kamenev, I. Dornic, G. \'Odor, F. de los Santos,
and H. Chat\'e for useful comments, and H. Hinrichsen for a very
helpful exchange of correspondence. We acknowledge financial support
from the Spanish MEyC-FEDER, project FIS2005-00791 and from Junta de
Andaluc{\'\i}a as group FQM-165.


\begin{thebibliography}{99}

\bibitem{EK} V. Elgart and A. Kamenev, Phys. Rev. E {\bf 74}, 041101 (2006).

\bibitem{MD}
        J. Marro and R. Dickman, {\em Nonequilibrium Phase Transitions
        in Lattice Models} (Cambridge University Press, Cambridge,
        1998).
               
\bibitem{Reviews_AS}
H. Hinrichsen, Adv. Phys. {\bf 49}, 815 (2000).  G. \'Odor,
Rev. Mod. Phys. {\bf 76}, 663 (2004).  G. Grinstein and
M. A. Mu{\~n}oz, in {\it Fourth Granada Lectures on Computational
Physics} edited by P.  L. Garrido and J. Marro, Lecture Notes in
Physics Vol 493 (Springer, Berlin), p. 223.

\bibitem{4} Additionally, they also identify some other classes
(not enumerated here) marginal in one dimension.

\bibitem{2species} Some of these universality classes
might have a better representation in terms of two species
\cite{Review_PCPD}. Also, some other classes with absorbing
states as, for example, the one of stochastic sandpiles
  [R. Dickman, M. A.  Mu{\~n}oz, A. Vespignani, and S. Zapperi,
    Braz. J. of Physics {\bf 30}, 27 (2000)]
usually represented by two density fields, admit a one-species
non-Markovian description. Hence, saying that there are $5$
non-trivial one-species absorbing-state universality classes is
somehow ambiguous.

\bibitem{conjecture}
        J.L. Cardy and R.L. Sugar, J. Phys. A  {\bf 13}, L423 (1980).
        H.K. Janssen, Z. Phys. B {\bf 42}, 151 (1981).
        P. Grassberger, Z. Phys. B {\bf 47}, 365 (1982).
                            
\bibitem{PC}
        P. Grassberger, F. Krause, and T. von der Twer, J. Phys. A
        {\bf 17}, L105 (1984). See also \cite{Reviews_AS} for reviews
        on this.


\bibitem{GV} O. Al Hammal, H. Chat\'e, I. Dornic, and M. A. Mu\~noz, 
Phys. Rev. Lett. {\bf 94}, 230601 (2005).


\bibitem{Review_PCPD} For an overview, see M. Henkel and H. Hinrichsen,
J. Phys. A {\bf 37}, R117 (2004).

\bibitem{support} Indeed, the results in \cite{EK} 
can be taken as a new evidence supporting the conclusion that PCPD
differs from DP.

\bibitem{TCPD}
G. \'Odor, Phys. Rev. E {\bf 73}, 047103 (2006).

\bibitem{Exact} M. Burschka, C. R Doering, and D. ben-Avraham, 
Phys. Rev. Lett. {\bf 63}, 700 (1989). 
 D. ben-Avraham,  M. Burschka, and C. R Doering,
J. Stat. Phys. {\bf 60}, 695 (1990).
 See also,
K. Krebs, M. Pfannmueller, B. Wehefritz, and H. Hinrichsen
J. Stat. Phys. {\bf 78}, 1429 (1995).
H. Hinrichsen, S. Sandow, and I. Peschel, 
J. Phys. A. {\bf 29}, 2643 (1996).
H. Hinrichsen, K. Krebs, and I. Peschel, 
Z. Phys. B {\bf 100}, 105 (1996). 
H. Hinrichsen, V. Rittenberg, and H. Simon,
 J. Stat. Phys. {\bf 86},
1203 (1997).




\bibitem{Empty} D. ben-Avraham, 
 Mod. Phys. Lett. B {\bf 9}, 895 (1995 ).  D. ben-Avraham, in {\it
 Statistical Mechanics in one-dimension}, Ed. V. Privman, pg. 29,
 Cambridge University Press, Cambridge 1997.



\bibitem{CT} J. Cardy and U. C. Ta\"uber, Stat. Phys. {\bf 90}, 1 (1998).

\bibitem{Jack} In R. Jack, P. Mayer, and P. Sollich, J. Stat. Mech. (2006) P03006, 
conclusions identical to ours are obtained for the {\it reversible
case} by exploiting the existence of {\it detailed balance}.

\bibitem{Peliti} L. Peliti, J. Phys. A  {\bf 19}, L365 (1986).
M. Droz and L. Sasvari, Phys. Rev. E {\bf 48}, R2343 (1993). See also,
B. P. Lee, J. Phys. A {\bf 27}, 2633 (1994).  B. P. Lee and J. Cardy,
J. Stat. Phys. {\bf 80}, 971 (1995).  J. Cardy, {\em Field Theory and
Nonequilibrium Statistical Mechanics}, (Troisi\`eme cycle de la
Physique en Suisse Romande, 1998-1999, semestre d'\'et\'e).
http://www-thphys.physics.ox.ac.uk/users/JohnCardy/home.html


\bibitem{RD} For a recent and nice review see, U. C. T\"auber, M. Howard, and
 B. P. Vollmayr-Lee, J. Phys. A. {\bf 38}, R79 (2005). For
 non-perturbative calculations see L. Canet, H. Chat\'e, and
 B. Delamotte, Phys. Rev. Lett. {\bf 92}, 255703 (2004). See also
 \cite{CT}.




\bibitem{GF}
M. Doi, J. Phys. A {\bf 9}, 1479 (1976).  L. Peliti, J. Physique, {\bf
46}, 1469 (1985).  P. Grassberger and M. Scheunert,
Fortschr. Phys. 28, 547 (1980).  C. J. DeDominicis, J. Physique
{\bf 37}, 247 (1976).  H. K. Janssen, Z. Phys. B {\bf 23}, 377 (1976).
P. C. Martin, E. D. Siggia, and H. A. Rose, Phys. Rev. A {\bf 8}, 423
(1978).

\bibitem{ZJ} J. Zinn-Justin, {\em
Quantum field theory and critical phenomena}, (Oxford University
Press, 4th edition, 2002).




  


\bibitem{Poisson} Alternatively, Eq. (\ref{action}) can also be obtained 
by employing the Poissonian transformation method \cite{Gardiner}. It
has been proved that both formalisms are completely equivalent:
M. Droz and A. McKane, J. Phys A: Math. Gen. {\bf 27}, L467 (1994)).


\bibitem{Gardiner}
C. W. Gardiner, {\it Handbook of Stochastic Methods}, Springer-Verlag,
Berlin and Heidelberg, 1985. M. A. Mu\~noz, Phys. Rev. E. {\bf 57},
1377 (1998).



\bibitem{MF} By imposing the first derivatives of the Hamiltonian, $H$, with
respect to $\phi$ and $\pi$ to vanish, one can identify the
homogeneous classical (mean-field) stationary solution. The
non-trivial expectation value of $\phi $ is $\langle \phi \rangle =
\mu/\sigma$ and hence, at mean-field level, the critical point is located at
$\mu_c=0$ and the order-parameter critical exponent is $\beta=1$.  At
this critical point the coefficients of both the linear-deterministic
and the leading noise terms vanish.



\bibitem{norenorm} As the diagrams in eq.(\ref{series}) are marginally
divergent in $d=2$, their derivatives do not require extra
renormalization, and therefore the fields do not have anomalous
dimensions \cite{Peliti}.


\bibitem{note1} Simple inspection of diagrams for different vertices, 
even those having the same topology (for example, the one-loop
diagrams contributing to DP \cite{conjecture}),
readily leads to the conclusion that combinatorial factors are
different, and therefore, each bare constant would have distinct
corrections.  

\bibitem{2loop} More complicated two-loop diagrams are generated when higher order 
vertices are generated, breaking the super-renormalizability of the
theory (see, for instance, fig. 3b in \cite{CT}).


\bibitem{nnn} Analogously, non-reversibility can also be included in models 
without a proper absorbing state as $2 A \leftrightarrow 0$ by
switching on reactions as $0 \rightarrow 3A, ~ 4A, ...$, without
affecting their critical behavior.

\end{thebibliography}
 \end{document}